\documentclass[preprint,aps,showpacs]{revtex4}
\usepackage{epsfig}
\begin{document}
\title{Glassy dynamics of Josephson arrays on a dice lattice}
\author{Vittorio Cataudella}
\affiliation{
	INFM $\&$ Dip. di Scienze Fisiche, Universit\'a ``Federico II",
	Complesso Universitario M.S. Angelo, 80126 Napoli, Italy}
\author{Rosario Fazio}
\affiliation{
	NEST-INFM $\&$ Scuola Normale Superiore, I-56126 Pisa, Italy}
\date{\today}

\begin{abstract}
In this Letter we study the phase diagram and the approach to the equilibrium 
of Josephson junction arrays on a dice lattice for two different values of 
the magnetic frustration ($f=1/2,1/3$). In both cases the array unedrgoes 
a Berezinskii-Kosterlitz-Thouless like phase transition. 
For the fully frustrated array, $f=1/2$, we find evidence 
of a {\em glassy behavior} although the system has {\em no disorder}. 
For comparison we study dynamics of the array at a different
value of frustration ($f=1/3$) and find no trace of glassy dynamics. 
\end{abstract}

\pacs{PACS numbers: 72.15, 73.23, 74.25}

\maketitle



Quantum interference can lead to localization even in the absence of any
disorder. This effect has been theoretically predicted~\cite{vidal} in tight
binding models on dice lattices, and experimentally verified in
superconducting wire networks~\cite{abilio} and semiconductor
heterostructures~\cite{naud}. This field-induced localization can be
understood in simple terms. An electron propagating through the lattice (to
be specific we consider the tight binding model of Ref.~\cite{vidal}), will
enclose Aharonov-Bohm loops. For specific values of the applied magnetic
field interference is destructive and the electron wavefunction is then
confined in the so called Aharonov-Bohm (AB) cages. In the case of
superconducting wire networks the existence of the AB cages manifests in a
suppression of the critical current and of the superconducting critical
temperature when the dice network is fully frustrated~\cite{abilio,footnote}.
Unexpectedly, decoration espriments~\cite{pannetier,serret} have shown that
those localization effects are accompanied by the absence of any regular 
pattern in the low temperature vortex configuration

Together with wire networks, Josephson Junction Arrays (JJA)~\cite{jja},
i.e.  networks of superconducting islands connected through
Josephson junctions, are ideal systems where to investigate subtle effects 
induced by the lattice structure and applied magnetic field 
are. In the last decade there has been a great amount of
work on the various aspects of the (thermo)dynamics of the XY model~\cite
{proceeds}, of which JJA are experimental realization. Among those works a
large body of activity has been devoted to the study of frustration induced
by a magnetic field applied perpendicularly to the array~\cite{frustration}.
As in the case of wire networks~\cite{abilio,pannetier,serret}, one expects 
that the superconducting transition temperature for JJA would signal the 
presence of localization. Contrary to wire networks however, phase 
fluctuations,
described by the two-dimensional XY model, will play a dominant role.

\begin{figure}
\begin{center}
\epsfig{figure=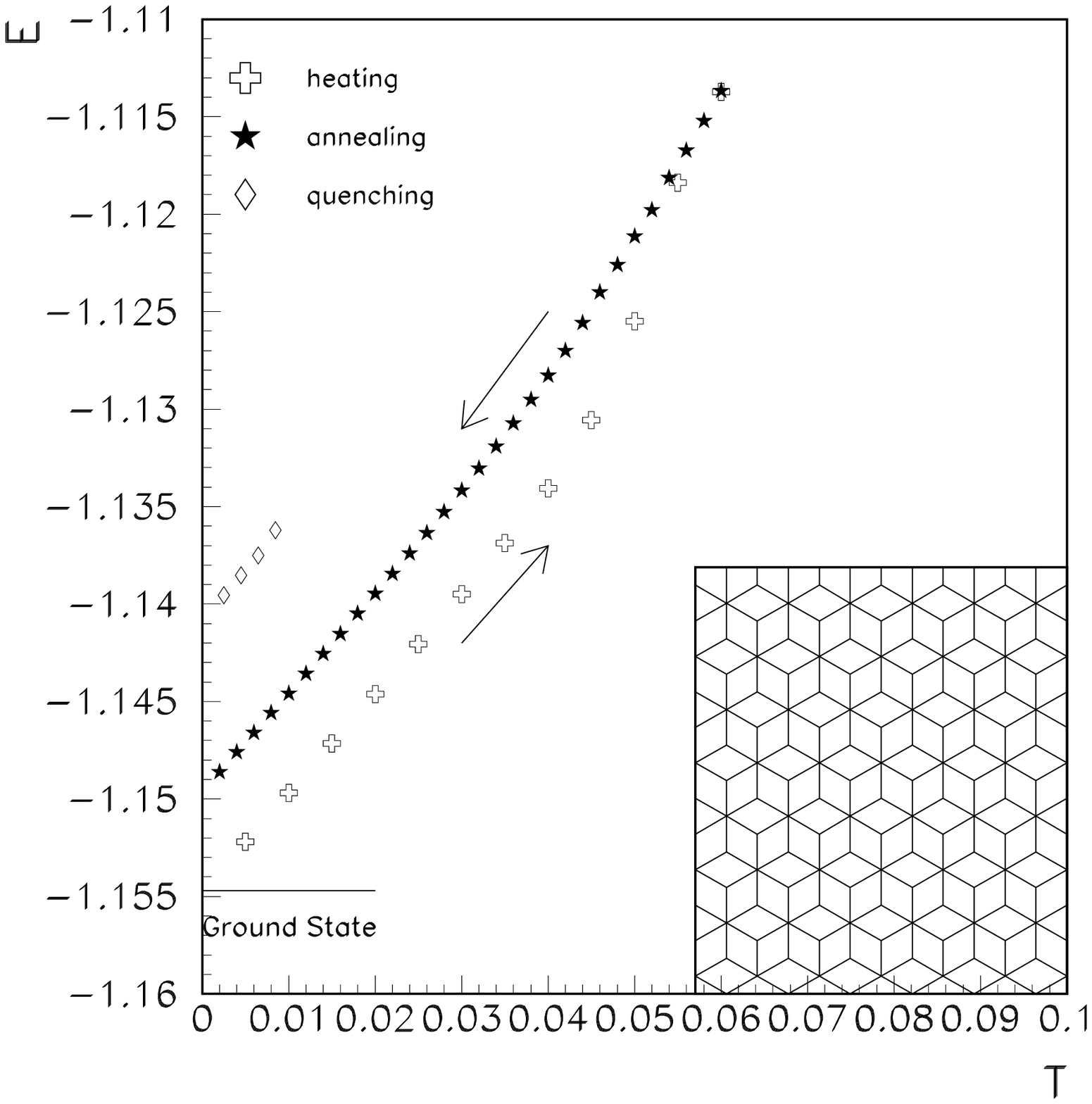,width=0.8\textwidth}
\end{center}
\caption{Ground state energy as a function of temperature in the 
fully frustrated case. The system size is $L=36$. The various 
curves are obtained
by heating (up arrow) and cooling (down arrow) the system . The third set of
data (with the largest energy) is obtained by quenching the system from very
high temperatures. In the inset it is shown the structure of the dice lattice.
Each link of the lattice is interrupted by a Josephson junction.}
\label{ground}
\end{figure}

Aim of this Letter is the study of the ground state and the finite temperature
properties of a fully frustrated XY model on a dice lattice, shown
schematically in the inset of Fig.~\ref{ground}, by means of Monte Carlo
simulations. Up to now the only theoretical prediction on this model has
been obtained by Korshunov~\cite{korshunov}. He has proposed the structure
of the ground state, both for wire networks and JJAs, and computed its 
degeneracy. His result is in apparent
contradiction with the experimental observation of the absence of ordering
in the vortex pattern at low temperature. Various explanations have been
proposed. The authors of Ref.~\cite{pannetier,serret} related the 
{\em unexpected disorder} in the vortex state with an infinite degeneracy and, 
in turn, with localization effects due to Aharonov-Bohm cages. On the other 
hand Korshunov's analysis~\cite{korshunov} shows that the family of 
ground states is ordered
and the absence of vortex regular pattern observed in the experiments could
be ascribed to that geometrical irregularities undoubtly present in the
experimental samples. What emerges from our work is that this apparent
discrepancy can be due to a glassy dynamics that sets in at low
temperatures. This result, we believe, can be of general interest since we
find a glassy behaviour in an  system {\em without}
any disorder. An investigation of JJA with a dice lattice can provide a
controlled way to investigate glassy states that are not consequence of
intrinsic random frustration~\cite{struik,halsey,aste}.

\begin{figure}
\begin{center}
\epsfig{figure=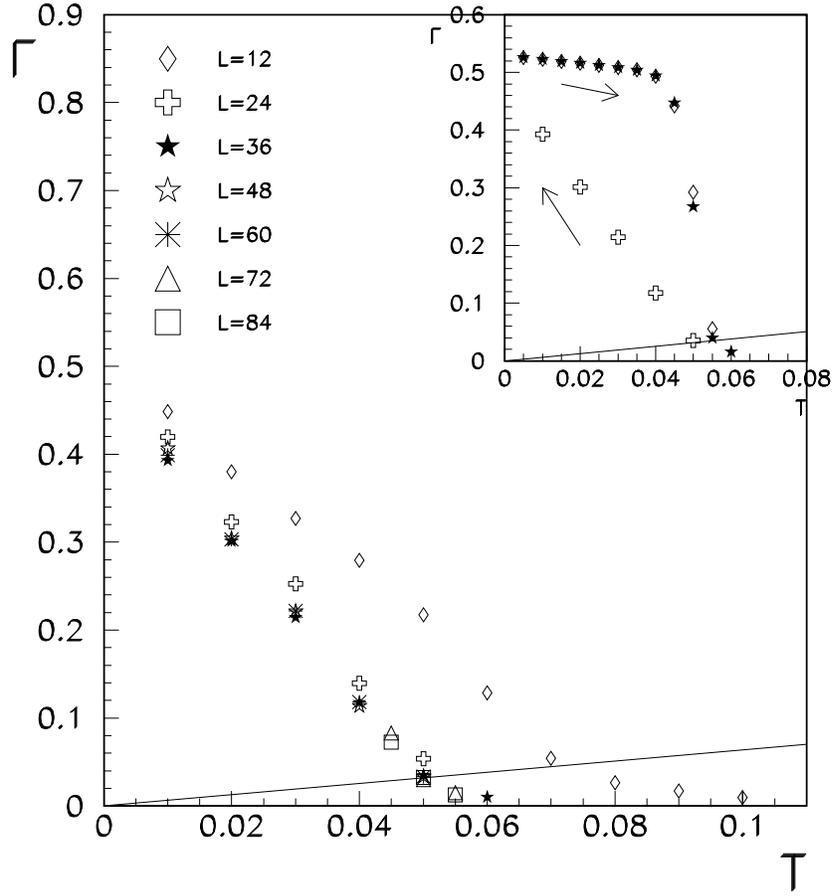,width=0.8\textwidth}
\end{center}
\caption{Temperature dependence of the stiffness for $f=1/2$. The different
symbols refer to different system sizes. In the inset we show the
hysteretic behaviour for the stiffness for system of size $L=36$. The arrows
indicate the heating and the cooling branches of the cycle.}
\label{rho}
\end{figure}

\begin{figure}
\begin{center}
\epsfig{figure=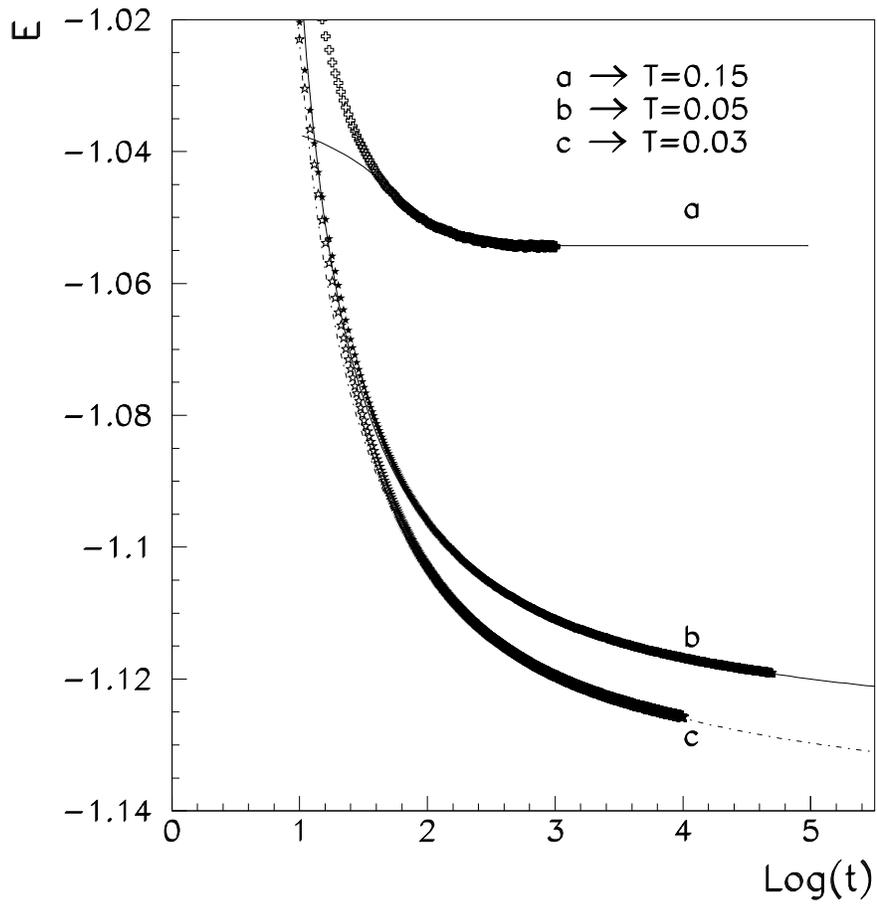,width=0.8\textwidth}
\end{center}
\caption{Energy relaxation as a function of the Monte Carlo time both in
disordered (curve a) and glassy (curves b,c) phases. In the disordered 
phase ($T=0.15$)
the decay is exponential. In the glassy phase the decay is much slower. 
The lines are fit to the data as discussed in the text. The time is 
measured in Monte Carlo steps.}
\label{energyrelax}
\end{figure}

\begin{figure}
\begin{center}
\epsfig{figure=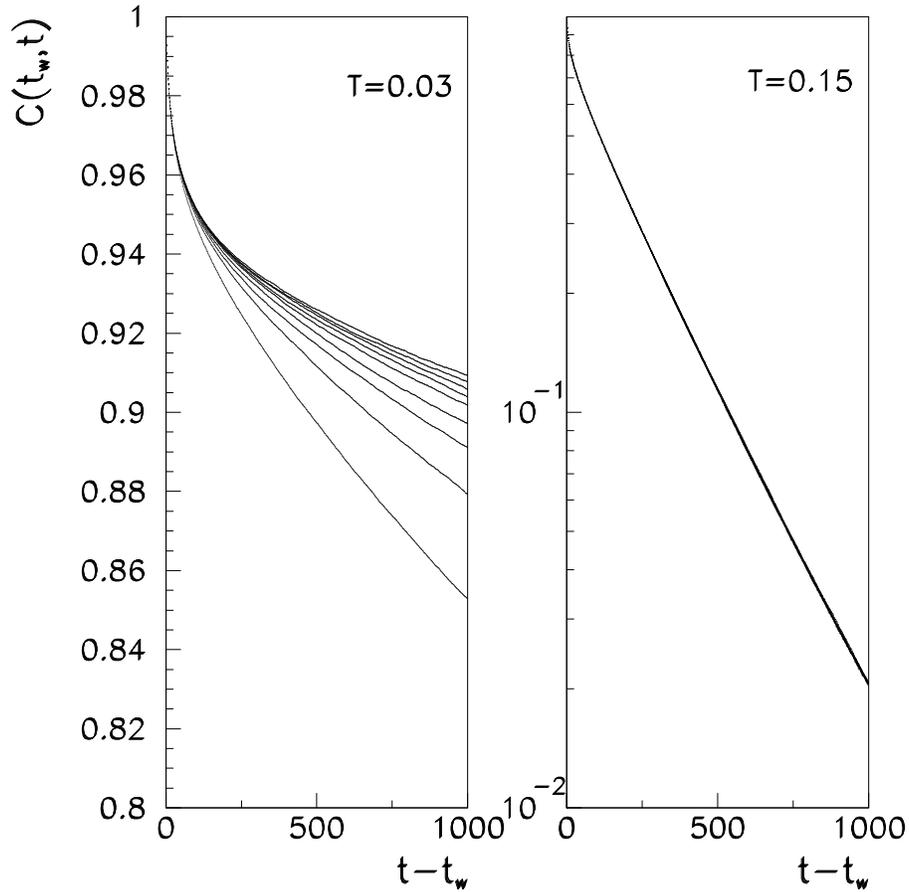,width=0.8\textwidth}
\end{center}
\caption{Two time correlation functions, for $f=1/2$, in the glassy state
and in the disordered phase.In the disordered phase ($T=0.15$) the decay
is exponential and there is no dependence on the waiting time. In the glassy
phase ($T=0.03$ upon cooling) the decay is much slower (see
the text) and there is a clear dependence on $t_w$. The different curves are
parameterized by $t_w=10^3k$ with $k=1-9$ from bottom to top. The time is
measured in Monte Carlo steps.}
\label{correlations}
\end{figure}

\begin{figure}
\begin{center}
\epsfig{figure=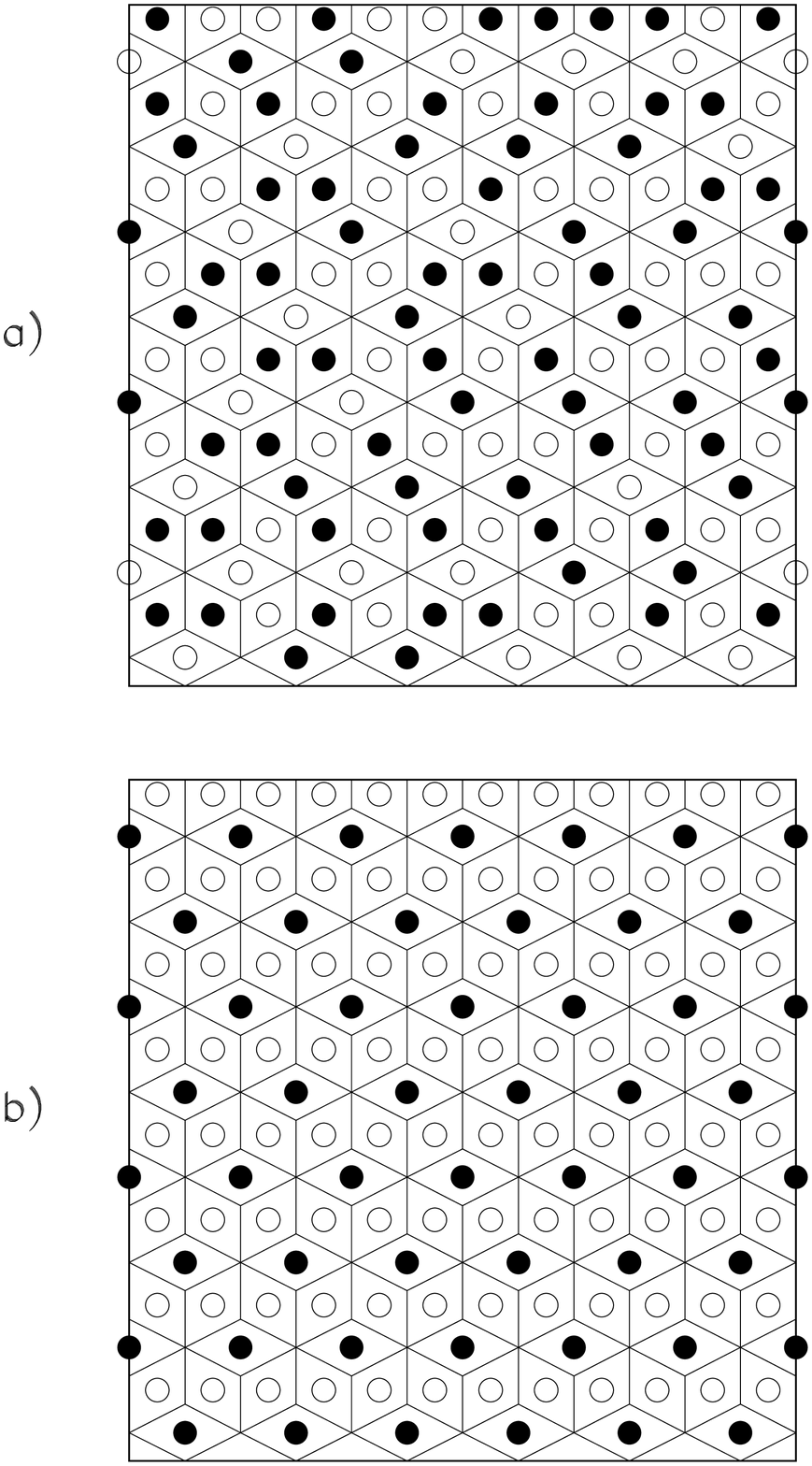,width=0.5\textwidth}
\end{center}
\caption{Typical vortex configurations, taken at low temperatures, for $f=1/2
$ (top) and $f=1/3$ (bottom) cases.}
\label{vortex}
\end{figure}

Our results are summarized in Figs.\ref{ground}-\ref{vortex}. We will show
that the JJA has an hysteretic behaviour
with a non exponential decay of phase correlations. 
In the last part of the Letter we reinforce the glassy behaviour found in
the fully frustrated case by comparing it with a JJA on a dice lattice with
a different value of the magnetic frustration.

The JJA is described by the XY Hamiltonian 
\begin{equation}
H=-E_{J}\sum_{<i,j>}\cos (\theta _{i}-\theta _{j}-A_{ij})  \label{model}
\end{equation}
where $E_{J}$ is the Josephson coupling and $<\ldots >$ refers to the sum
over nearest neighbors. In the experimental situation $E_{J}$ changes with
temperature, here for simplicity we treat it as an independent parameter.
The variables $\theta _{i}$ are the phases of the
superconducting order parameter of the i-th island. 
For later convenience we introduce the phase difference $\theta _{ij}=\theta
_{i}-\theta _{j}$. The magnetic field
enters through $A_{ij}=(2\pi /\Phi _{0})\int_{i}^{j}{\bf A}{\bf \cdot }d{\bf %
l}$ (${\bf A}$ is the vector potential), where $\Phi _{0}=hc/2e$ is the flux
quantum. The relevant parameter which describes the magnetic frustration is $%
f=\sum A_{ij}/(2\pi )$, where the summation runs over the elementary
plaquette. We study the cases $f=1/2$ and $f=1/3$ on a dice lattice.
Estimation of the various quantities have been obtained averaging at least $%
10^{7}\cdot L^{2}$ Monte Carlo configurations by using a standard Metropolis
algorithm. The largest lattice studied is $L=84$. We refer to the Monte
Carlo dynamics as to the dynamics of the system. Energies and temperatures 
will be expressed in units of $E_J$ and $E_J/k_B$ ($k_B$ the Boltzman 
constant) respectively.

\noindent \underline{f=1/2.} The ground states for the fully
frustrated case have been proposed by Korshunov~\cite{korshunov}. We have
tested his conjecture performing simulations at very low temperatures
starting from one of the ground states proposed. Our simulations give a
value for the ground state energy (per site) $E_{G}(f=1/2)=-1.155$ in
very good agreement with the analytic calculations (see Fig. \ref{ground}).
The stiffness 
\[
\Gamma =\frac{\partial ^{2}{\cal F}}{\partial \delta ^{2}}\;\;,
\]
used to signal the presence of the transition, is defined through the
increase of the free energy ${\cal F}$ due to a phase twist $\delta $
imposed in one direction~\cite{ohta}. Our data, plotted in Fig.\ref{rho} 
for $\Gamma $ indicate that
there is a phase transition. The critical temperature was determined by
using the following {\em ansatz} for the size dependence of $\Gamma $~\cite
{weber} 
\[
\frac{\pi \Gamma }{2T_{c}}=1+\frac{1}{2\ln (L/l_{0})}
\]
where $l_{0}$ is the only fit parameter. The estimated value is 
$T_{c}=0.05$.
Note that the transition temperature for the fully frustrated square lattice
($T_c\simeq 0.446$) is one order of magnitude larger than the one found for
the dice lattice.
This type of scaling implies that the transition belongs to the
Berezinskii-Kosterlitz-Thouless universality class~\cite{BKT}. Although larger
systems are needed to confirm this point, however the very existence of the
critical point is clearly seen in the simulations.

So far we described the thermodynamics of the model introduced in Eq.(\ref
{model}). A closer inspection to the approach of the system to the
equilibrium reveals an interesting scenario. In Fig.\ref{ground} the energy
per site, obtained by heating and cooling the system, is shown as a 
function of temperature. 
The low energy curve is obtained by starting at very low
temperatures in one of the ground states and heating up the system. A clear
hysteretic behaviour is observed up to temperatures of the order of 
$0.06$. Upon cooling, the energy of the system remains higher and never 
reaches the ground state value. Finally the top-most symbols ($\diamond$) 
in Fig.\ref{ground} represent the average internal energy in the 
case in which the system is quenched from a very high temperature state. 
Further confirmation of the glassy behaviour is
seen in the behaviour of the stiffness upon cooling and heating, see the
inset of Fig.\ref{rho}. As expected, also $\Gamma $ shows hysteresis and
upon cooling it is much lower than in the heating part of the cycle implying
that the system is not able to order. The transition temperature,
see the discussion in the previous paragraph, was determined
considering the data obtained upon cooling the sample. Essentially the same
temperature is obtained scaling the data on the heating run. The transition
at which the hysteretic behaviour sets in appears to be higher than the
transition to the ordered state. On cooling down from higher temperatures,
the onset of glassy behaviour will prevent the system to enter in the
ordered state. This is also the main reason why it was necessary to consider
systems as large as $L=84$ to ascertain the very existence of the phase
transition.

The presence of a glassy behaviour should manifest in an anomalous
relaxation to the equilibrium state. We analyzed how the energy shown in Fig.%
\ref{ground} relax to the stationary value. An accurate fitting of the
energy decay with time show that for temperatures $T\leq 0.06$ the system
enters a glassy state characterized by a slow (logarithmic) relaxation to
the equilibrium situation. The fits reported in Fig.\ref{energyrelax} show
the change from an exponential behavior at high
temperatures to a decay typical of glassy dynamics 
\begin{equation}
E(t)=a+\frac{b}{\log(t/\tau )}
\label{erelax}
\end{equation}
for low temperatures ($a,b,\tau$ fitting parameters). 
The equilibrium value estimated from the logarithmic 
fit of Eq.(\ref{erelax}) is consistent with the value obtained in the 
heating branch shown in Fig.\ref{ground}

In the presence of slow relaxation, a more detailed analysis on the approach
to the equilibrium can be made by studying the correlation 
\begin{equation}
C(t_{w},t)=\frac{1}{L^{2}}\langle \mid \sum_{ij}\cos (i[\theta
_{ij}(t)-\theta _{ij}(t_{w})])\mid \rangle 
\end{equation}
where $t_{w}$ is the waiting time (i.e. the time that the system is kept at
that temperature before the measurement starts) and $t$ is the time. In Fig.%
\ref{correlations} the results at low and high temperature are presented. At
high temperature, ($T=0.15$ in the figure), the correlation $C(t_{w},t)$
decays exponentially as a function of $t-t_{w}$ and it is independent
on $t_{w}$. More interestingly at temperatures below $T=0.06$ the loss
of correlations is much slower and it depends strongly on $t_{w}$. 

The evidences presented so far indicate that the presence of AB cages not only 
affects the $T_c$ or the critical current but induce a glassy behavior in the 
low temperature region where phase fluctuations play a dominant role.
In order to strengthen this conjecture we analyzed the properties of JJA on 
a dice lattice at $f=1/3$. For this value of frustration no localization is
predicted.

\noindent \underline{f=1/3.} The ground state, in this case, has an energy
per site $E_{G}=-4/3$. Differently from the fully frustrated case, a
study of the energy as a function of the heating/cooling cycles did not show
any hysteretic behaviour. The study of stiffness as a function of
temperature, signals a phase transition at a temperature of the order of $%
T_{c}\sim 0.2$, a much higher value than for the fully frustrated case.
What is most important is that in the case of $f=1/3$ there is not trace of
hysteretic behaviour neither in energy nor in the stiffness and the
correlation function $C(t_{w},t)$ decays exponentially with the time.

Finally we comment on the experimental issue related to the presence of
glassy dynamics related to the AB cages. In the experiments reported in
Refs.\cite{pannetier,serret}, decoration measurements show a regular vortex
pattern in the $f=1/3$ case while there is no trace of ordering for $f=\frac{%
1}{2}$. In Fig.\ref{vortex} we make the same type of comparison. It is
evident that in fully frustrated case there is no trace of the vortex
lattice, although there is an ordered phase at low temperature as confirmed
by the analytic approach of Ref.\cite{korshunov} and the simulations of the
present work. The vortex configurations seems to reproduce the experimental,
however some words of cautions are necessary. The experiments were performed
in wire networks while in this paper we studied JJAs. At low temperatures the
two systems have a different energy dependence on the phase difference 
($\cos (\theta )$ for arrays and $\theta^{2}$ for wires) as discussed 
in Ref.\cite{korshunov}. The phase configurations, however, do
not differ significatively in the two cases, and this makes us confident on
the interpretation.
Nevertheless it would be important to have experimental
results on JJA to confirm or dispute this analysis. 

A final point that
should be made is related to the dynamics. Here we discussed a Monte Carlo
dynamics. In principle the existence of a glassy behaviour can depend on the
choice of the dynamics. It would be important to test the ideas put forward
in this Letter by performing simulations of the corresponding RSJ~\cite{kim}
and local damping~\cite{korshunov2} dynamics which gave in the past very
good agreement with experiments. This is outside the scope of this work.
However we emphasize once more that the evidence of the apparent loss of a
regular pattern in the vortex configuration seen in the experiments fits
nicely with the phenomenology found in the present work.

The authors would like to thank E. Serret and B. Pannetier for useful
discussions and S.E. Korshunov for a critical reading of the manuscript. 
This work was supported by the INFM and by and the European
Community (Contract FMRX-CT-97-0143).

\end{document}